\documentstyle[multicol,aps,psfig]{revtex} 
\begin{document} 
 
\draft 
\preprint{LBL-45309} 
 
\title{Energy and Centrality Dependence of Rapidity Densities 
at RHIC} 
 
\author{Xin-Nian Wang} 
\address{Nuclear Science Division, Mailstop 70-319, 
        Lawrence Berkeley National Laboratory, Berkeley, California 94720.} 
\author{Miklos Gyulassy} 
\address{Department of Physics, Columbia University, 538 W. 120th St.,
  New York, NY 10027} 
\date{August 8, 2000} 
\maketitle 
\begin{abstract} 
The energy and centrality dependence of the charged multiplicity per  
participant nucleon is shown to  be able to differentiate between final-state
saturation and fixed scale pQCD models of initial entropy production in 
high-energy heavy-ion collisions. The energy dependence is shown to 
test the nuclear enhancement of the mini-jet component  
of the initial conditions, while the centrality dependence provides a key  
test of whether gluon saturation is reached at RHIC energies. 
HIJING model predicts that the rapidity density per participant
increases with centrality, while the saturation model prediction 
is essentially independent of centrality. 
\end{abstract} 
 
 
\begin{multicols}{2}  
   
 Signals of the formation of a quark-gluon plasma in high-energy 
 heavy-ion collisions are sensitive to the initial condition of the 
 dense matter created in the early stage of the collisions.  
 Global bulk observables at the Relativistic Heavy-ion Collider (RHIC)
 such as the rapidity density of
 multiplicity and 
  transverse energy  provide important constraints on 
  those initial conditions. The first data on $dN_{ch}/d\eta$ in Au+Au 
  reactions at $\sqrt{s}=65$ and 130 AGeV were reported in Ref.~\cite{phobos}. 
  We focus here on how the systematics (energy and centrality 
  dependence) of this observable can be used to differentiate between 
  competing entropy production mechanisms.  The entropy is especially 
  interesting because it remains approximately conserved during the 
  complex dynamical evolution of the system {\em if} local equilibrium is
  maintained.  In longitudinal boost invariant hydrodynamics the 
  rapidity density of particles is in fact conserved.  

Theoretical estimates of 
  the initial condition at RHIC vary over a wide range\cite{qm99} because of 
  the (as yet) unknown interplay between soft and hard mechanisms of 
  multiparticle production in nuclear collisions.   
Phenomenological models ranging from soft string models \cite{fritiof,urqmd},  
perturbative QCD (pQCD) models \cite{hij135,vni,nexus,bm,klf,EKRT},  
and classical Yang-Mills approaches \cite{mv} have been proposed to 
predict the produced entropy.  In these models there are basic
physical parameters, such as the nuclear size dependence 
of the scale, $p_0(\sqrt{s},A)$, separating perturbative QCD 
and nonperturbative components, that control the magnitude 
of entropy production. In multiple mini-jet pQCD-inspired models, 
such as HIJING, the unknown nuclear gluon shadowing and parton 
energy loss due to final state interactions also 
lead to a factor of $\sim 2$ variation in the predictions for 
the final charged hadron rapidity density in the central region \cite{qm99}. 
With the first measurements of the $dN_{ch}/d\eta$ from 
the PHOBOS experiment\cite{phobos} and the other experiments 
at the Relativistic Heavy-ion Collider (RHIC), 
the theoretical uncertainties can soon be considerably reduced.  
 
The measured $dN_{ch}/d\eta (|\eta|<1)$ for central collisions  
was reported to agree within statistical and systematic errors with 
the default HIJING1.35 \cite{hij135} predictions.  Of course, one 
cannot conclude that the dilute HIJING1.35 initial conditions (with 
$dN_{glue}(p_T >2\;{\rm GeV})/dy\approx 250$ at $\sqrt{s}=130$ AGeV)  
are correct without much more extensive differential experimental  
studies, especially of the shape of $dN_{ch}/d\eta$ and the  
high $p_T$ hadron spectra \cite{wgprl,wang98}.
HIJING does not include final state rescattering  
except for schematic jet quenching. Other models, such as 
in EKRT\cite{EKRT}, that assume local equilibrium and hydrodynamic 
expansion can also reproduce the first data starting from much higher  
initial conditions. The aim of this letter is to emphasize that 
the centrality or impact parameter dependence of 
the charged particle rapidity density provides a key observable that, 
combined with other differential measurements, can significantly 
narrow the current band of uncertainty of the initial conditions 
produced at RHIC and search for evidence of novel gluon saturation 
phenomena\cite{bm,EKRT,mv}  or dynamical screening 
effects\cite{emw} at high density. 
 
In this study, we use  HIJING Monte Carlo model version 1.35. This model  
incorporates pQCD to compute multiple mini-jet production and  
uses the Lund string model\cite{fritiof} to describe 
soft beam jet fragmentation  
and hadronization of the jets. In default HIJING1.35, nuclear shadowing of   
gluons is assumed to be identical to the observed quark shadowing.
The large $p_T$ quarks are assumed to have energy 
loss of $dE/dx=1$ GeV/fm, and jet quenching is modeled by a  gluon 
splitting scheme. HIJING unitarizes the minijet cross section  
via an eikonal approach\cite{wang91}
\begin{equation} 
\sigma_{in}^{NN}=\int d^2b 
[1-e^{-(\sigma_{soft}(s)+\sigma_{jet}(s))T_{NN}(s,b)}], 
\end{equation}
where $\sigma_{jet}(s)$ is the inclusive jet production cross
section with $p_T>p_0$, $\sigma_{soft}$ accounts for soft interaction cross
section and $T_{NN}(s,b)$ is the nucleon-nucleon geometrical overlap function.
The two critical physical parameters of this model, $p_0$ and $\sigma_{soft}$,
are adjusted to fit the measured cross sections and $dN_{ch}/d\eta$ for
inelastic $pp(\bar{p})$ collisions at high energies.  In versions 1.35 and
below Duke-Owen (DO) \cite{DO} parameterization of parton distributions is
employed.
If more recent parametrization of parton distributions in nucleons is
used, one has to use an energy-dependent $p_0$ in order to fit the cross 
section and hadron rapidity density in $pp$ or $p\bar{p}$. However, the
results will remain the same as presented in this letter using HIJING1.35.
The important point to keep in mind is
that the cut-off scale $p_0$ is fixed in HIJING by $pp(\bar{p})$ data and is
assumed to be {\em independent} of $A$ or the centrality.

Figure~1 compares  $dN_{ch}/d\eta$ per pair of participant 
nucleons in $pp(p\bar{p})$ and central $Au+Au$ collisions as  
functions of colliding 
energy. For nuclear collisions the two solid curves correspond to 
HIJING1.35 calculations with (upper) and without (lower) jet quenching.  
We note that the effect of jet quenching on the total hadronic multiplicity  
in HIJING model only becomes significant above $\sqrt{s}>100$ GeV.  
 
From  Fig.~1 both $pp(\bar{p})$ and $AA$ collisions 
appear to have  a  component that is  approximately constant 
$\langle n\rangle_{soft}\approx 1.3$ 
plus a logarithmic energy dependent component.  
The constant component arises 
from the  soft (low transverse momentum) hadron production 
due to  beam jet string fragmentation in HIJING. 
This  soft particle production 
is also proportional to the number of participants, and therefore 
its contribution to $dN_{ch}/d\eta$ per 
participant is {\em independent} of $A$ in heavy-ion collisions. 
Including the contribution from minijet production on the other
hand, the total hadronic rapidity density acquires the form
\begin{equation}
\frac{dN_{ch}}{d\eta}=\langle N_{part}\rangle \langle n\rangle_{soft} 
+ f\;\langle N_{binary}\rangle \frac{\sigma_{jet}^{AA}(s)}{\sigma_{in}^{NN}}
 \label{eq:nch} 
\end{equation}
where $\sigma_{jet}^{AA}(s)$ is the averaged inclusive jet cross 
section per $NN$ collision in $AA$ collisions and $f\approx 1.2$. 
We have checked that HIJING results without jet quenching
indeed follows the above scaling. Since the default HIJING includes
nuclear shadowing of the gluon distribution, $\sigma_{jet}^{AA}$ 
is generally  smaller than $\sigma_{jet}^{NN}$. 
For central $Au+Au(b=0-3{\rm fm})$ collisions the averaged number of 
participants is $\langle N_{part}\rangle\approx 350$ and the 
averaged number of binary collisions per participant pair is 
$2\langle N_{binary}\rangle/\langle N_{part}\rangle \approx 4.7$ 
at RHIC energies. Because of the $s$ dependence on the minijet 
production cross section $\sigma_{jet}^{AA}(s)$, the energy dependence 
of the hadronic multiplicity is amplified by $\sim A^{1/3}$ in heavy-ion 
collisions relative to $pp(\bar{p})$ by the binary nature of semihard 
processes (see also Fig.~3 below). The rise of the multiplicity  
per participant relative to Super Proton Synchrotron (SPS)
is consistent, within the large 
(mostly systematic) error bars, with the predicted binary scaling 
of the hard component from $pp$ to $AA$ via  Eq.~(\ref{eq:nch}).
\begin{figure} 
\centerline{\psfig{figure=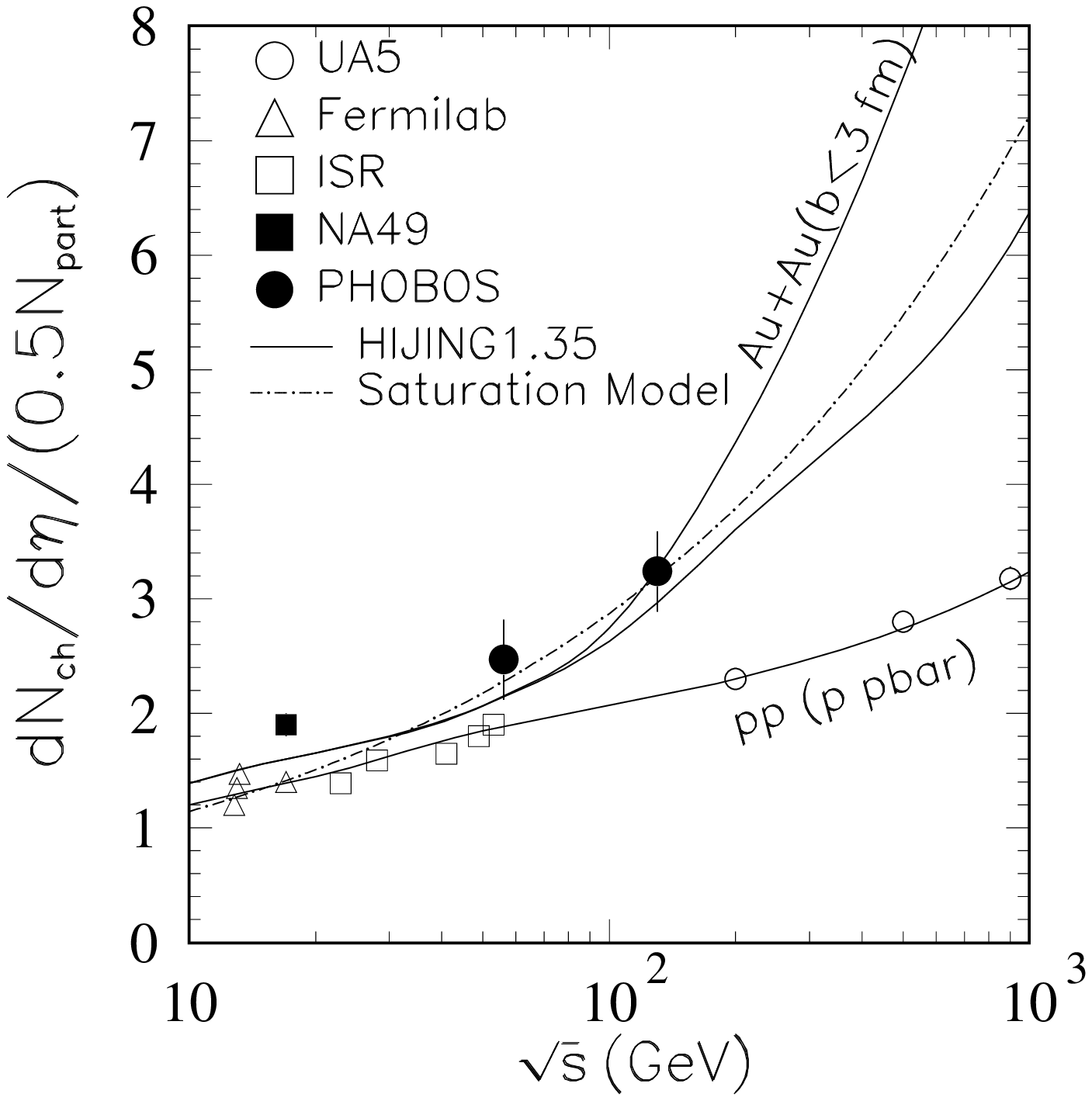,width=2.5in,height=2.5in}}
\begin{center}
\begin{minipage}[t]{8.6cm}
Charged particle rapidity density {\em per  participating 
baryon pair} versus the c.m. energy. The  PHOBOS 
data\protect\cite{phobos} (filled circles) 
for the  6\% most central Au+Au are  
compared to $pp$ and $p\bar{p}$ data (open symbols)  
\protect\cite{ppbar,ppisr,ppfermilab} 
and the NA49 $Pb+Pb$(central 5\%) data \protect\cite{na49} (filled square).  
HIJING1.35 (solid) with (upper) and without jet quenching (lower) 
and EKRT (dot-dashed) predictions are also shown.
\end{minipage}
\end{center}
\end{figure}

Parton saturation can occur in both initial and final state interactions but 
they both give the same kind of $A$-dependence of particle production. 
The initial state saturation model \cite{bm,mv} is based on the nonlinear 
Yang-Mills field effect arising from $gg\rightarrow g$, in which case the 
saturation scale is determined by $p_{sat}^2R^2=(9/16)C_A\pi\alpha_s\, AxG$ 
where $C_A=3$, $\alpha_s$ is the strong coupling constant and
$xG$ is the gluon distribution per nucleon at $x=2p_{sat}/\sqrt{s}$ 
and $Q=p_{sat}$. Numerical solution of the Yang-Mills field \cite{mv} could 
also include final state saturation and dynamical screening effects\cite{emw}.
In this paper we will only compare our calculation with EKRT\cite{EKRT} 
model of final state saturation which gives a definite prediction of 
both energy and $A$ dependence of rapidity density of charged multiplicity.  
In this model, the pQCD growth of low $p_T$ gluons is cut off below a 
saturation scale, 
$p_0(\sqrt{s},A)\equiv p_{sat}\approx 0.2 A^{0.13}(\surd s)^{0.19}$ 
at which $dN_g/dy=p_{sat}^2R^2$.

Assuming direct proportionality between parton 
and the final hadron number, the energy and atomic number dependence of 
the total hadronic multiplicity per unit rapidity in central $A+A$ 
collisions is estimated in EKRT \cite{EKRT} as 
\begin{equation} 
\frac{dN_{ch}}{dy}(b=0)\approx \frac{2}{3}1.16 A^{0.92}(\sqrt{s})^{0.4}. 
\label{eq:sat} 
\end{equation}
To compare to HIJING calculation and the PHOBOS data for pseudo-rapidity 
density, we scale the above result by a factor 0.9.
As shown in Fig.~1, the EKRT model estimate of the energy 
dependence of the entropy density is remarkably close to the 
HIJING results and is also consistent with the PHOBOS data.

However, a critical difference between these models
is that the hadronic multiplicity per participant 
actually {\em decreases} in the EKRT model with the atomic number of the  
colliding nuclei due to the saturation requirement. 
It therefore considerably overestimates the hadronic multiplicity 
if extrapolated down to  $pp(\bar{p})$.  
This is in contrast to the HIJING model of minijet production without  
saturation, where the multiplicity per participant increases with $A$ 
according to Eq.(\ref{eq:nch}).

\begin{figure} 
\centerline{\psfig{figure=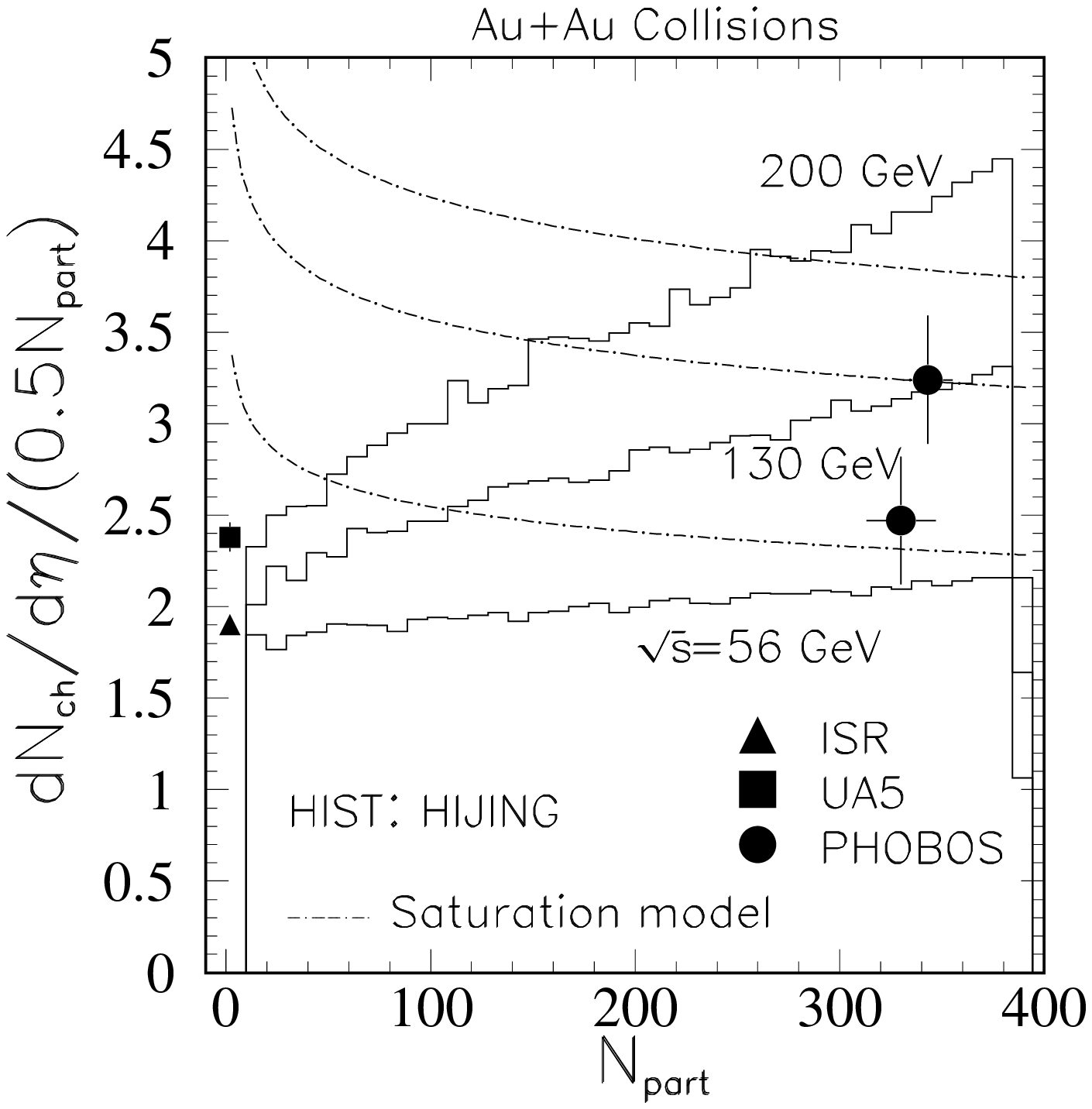,width=2.5in,height=2.5in}}
 \begin{center}
\begin{minipage}[t]{8.6cm}
The  charged particle (psuedo)
rapidity density {\em per participating baryon pair} is shown as  
functions of the number of participant baryons for  central 
$Au+Au (b<3{\rm fm})$ collisions at $\sqrt{s}=56$, 130 and 200 AGeV. 
Results of HIJING1.35 (solid histograms) are compared to $pp(\bar{p}$  
and PHOBOS data and to EKRT predictions (dot-dashed).
\end{minipage}
\end{center}
\end{figure}  
 
Since extrapolating the EKRT model down to $pp(\bar{p})$ collisions 
cannot be justified, it is more useful to study centrality dependence 
of heavy-ion collisions. We calculate next the centrality dependence of 
the hadronic multiplicity per participant in order 
to emphasize the difference between fixed scale and 
saturation models of entropy production.  Shown in Fig.~2 are 
the $dN_{ch}/d\eta$ per participant pair as  
functions of  $\langle N_{part}\rangle$ 
at three different energies together with the RHIC data by PHOBOS 
experiment \cite{phobos} and the $pp(\bar{p})$ data. The HIJING results  
increase monotonically with the number of binary collisions per 
participant $\langle N_{binary}\rangle/\langle N_{part}\rangle$ 
in an intuitive way as given in Eq.(\ref{eq:nch}). The slope
increases with energy since the hadronic multiplicity is proportional 
to jet cross section which has a significant energy dependence. For very 
peripheral collisions, the results agree with $pp(\bar{p})$ data because 
that is how the model parameters of HIJING are constrained. Naively,  
$\langle N_{binary}\rangle/\langle N_{part}\rangle 
\sim \langle N_{part}\rangle^{1/3}$. The deviation from such a simple 
dependence in HIJING calculation is due to a combined effect of 
jet quenching and dilute edges in Wood-Saxon nuclear distributions used 
in HIJING.  
 
The dot-dashed lines are the predictions extrapolated from
EKRT saturation model Eq.~(\ref{eq:sat}), assuming the dependence on 
atomic number in central collisions roughly the same as the number of 
participant pairs $N_{part}/2$ for a fixed $A$.
A generalization of the EKRT approach to model local 
saturation is expected not to change our estimate of its 
centrality dependence qualitatively.
Contrary to HIJING predictions, the saturation 
model gives increasing multiplicity per participant toward more peripheral 
collisions. While the  extrapolation  to the highest impact parameter 
(low participant number) domain is dubious, a general feature 
of saturation models is expected to be a weakly decreasing or
constant dependence on centrality in semi-peripheral to central collisions.  
The upcoming experimental data should easily distinguish these two 
widely different predictions.  
 
We emphasize the power of the combined 
energy and centrality dependence in Fig.~2 to search for novel 
nuclear saturation effects. This also applies at the Large Hadron 
Collider (LHC) energies. 
The higher the energy the greater the difference between the  
predicted centrality dependencies. 
A threshold for saturation can be identified experimentally
by looking for a region with negative derivatives 
$d(dN_{ch}/d\eta N_{part})/d N_{part}\le 0$ above some
$N_{part}>N_{crit}(\sqrt{s})$. 
Ordinary dynamical screening effects on the other hand, 
could be recognized by a gradual reduction of the positive  
derivatives predicted by HIJING with increasing centrality. 
 
Of course more differential data will eventually  provide much stronger 
constraints on models. Measurements of large $p_T$ hadron 
spectra can provide for example information 
related to jet quenching \cite{wgprl,wang98}. 
Another important  differential  observable is the rapidity dependence of  
$dN_{ch}/d\eta/\langle N_{part}\rangle$. The rapidity  
dependence reflects the $x$-dependence of the gluon distribution  
function in nuclei and should be strikingly different
in saturation models. In Fig.~3, the rapidity dependence of  
this quantity is shown for the HIJING model. Since most of the 
gluon production is from small $x$ region, one can see that the 
nuclear enhancement in the hadronic rapidity density per 
participant is mainly restricted to the central rapidity region
in this model. The width of this region is roughly determined by 
the scale $p_0$ as $\Delta y\sim \ln(\sqrt{s}/p_0)$. 
If saturation occurs, a different rapidity dependence of 
the nuclear enhancement may result. The nuclear size and energy 
dependence of the shape of that distribution as compared to different 
model calculations adds an important observable in the search for 
possible saturation effects

Finally we note that the study of the centrality dependence
of nuclear enhancement per participant
depends on the experimental ability to measure or deduce
$N_{part}$. At present this is done assuming that $dN_{ch}/d\eta\propto
N_{part}$ in the fragmentation regions $\eta>3$. In Fig.~3,
we see that at least in the HIJING model the nonlinear nuclear dependence
is indeed confined to $\eta<3$. However,
detailed studies of the centrality dependence of correlations between
 $dN_{ch}/d\eta$ at $\eta=0$ and $\eta>0$ could help 
reduce systematic errors.
 
\begin{figure} 
\centerline{\psfig{figure=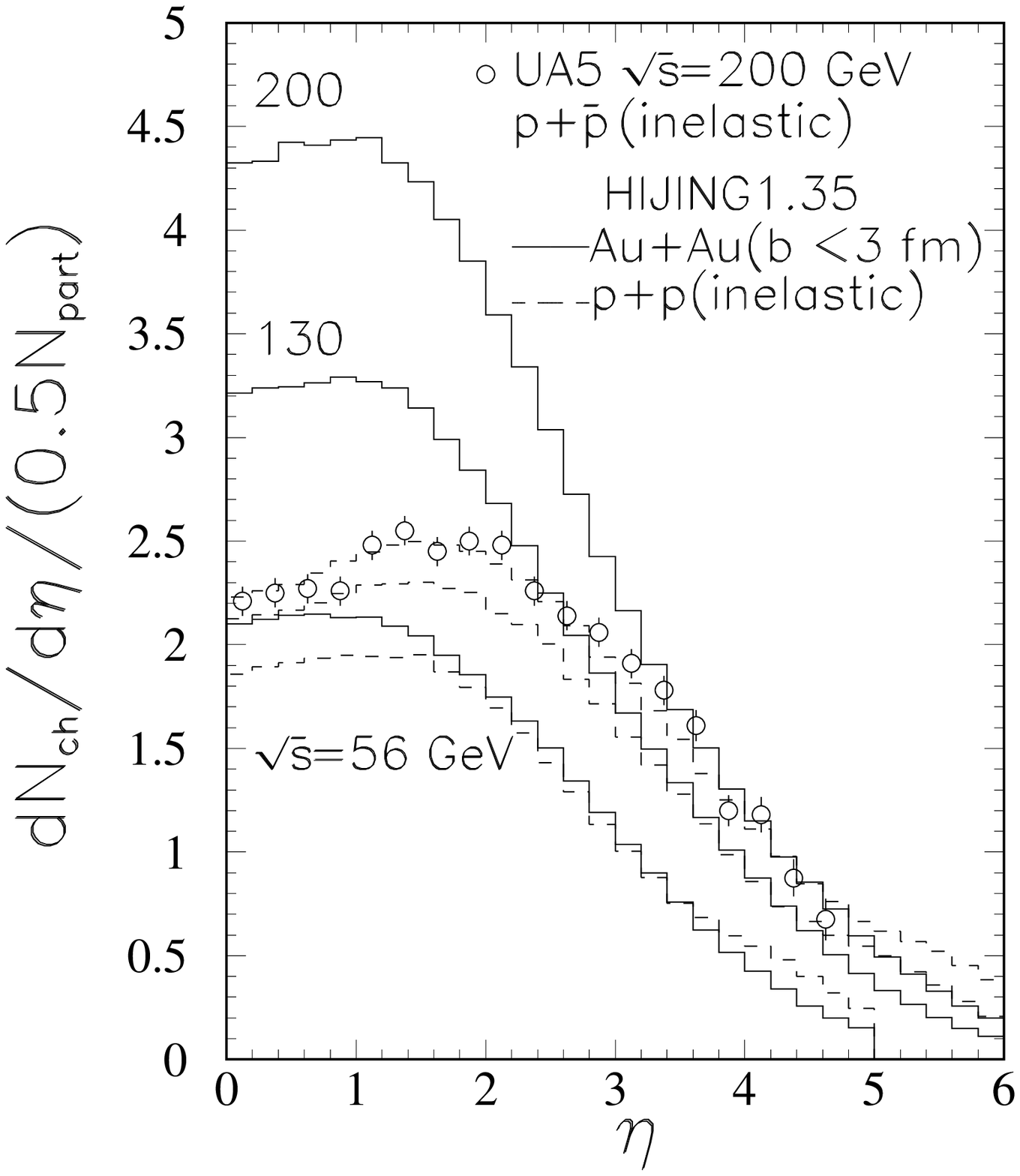,width=2.5in,height=2.5in}}
 \begin{center}
\begin{minipage}[t]{8.6cm}
The charged particle rapidity density {\em per participating 
baryon pair} as functions of rapidity as predicted by HIJING for central 
$Au+Au (b<3{\rm fm})$ (solid) and $pp$ (dashed) collisions  
at $\sqrt{s}=56$, 130 and 200 AGeV. Also shown are experimental data for  
$p\bar{p}$ collisions at $\sqrt{s}=200$ GeV.
\end{minipage}
\end{center}
\end{figure}  
 
In summary, we have shown how the energy and centrality dependence of the 
hadronic multiplicity densities in the central region of high-energy heavy-ion 
collisions can be used to constrain
the mechanisms responsible for producing the  initial 
conditions in such reactions. The energy dependence of  
the charged hadron multiplicity has  been shown by PHOBOS to be  
enhanced relative to $pp$ reactions and  is consistent 
with the onset of pQCD dynamics driven by 
the binary nature of semi-hard (mini-jet) processes. 
Though similar energy dependence is predicted by very different pQCD 
based models, the centrality dependence can differentiate
between them. In principle, parton saturation 
is expected to occur\cite{bm,EKRT,mv}  at asymptotic 
 high-energy in collisions of very heavy nuclei. The interesting question 
is where this occurs in practice. 
We proposed that the centrality or $A$ dependence
of the hadron multiplicity at RHIC and higher energies may help to 
answer this question.

{\em Note added in proof}: Since the submission of this letter, 
the first data \cite{phenix} on the centrality dependence of the 
multiplicity has become available. Other calculations, motivated by 
this work, have also appeared, e.g.,\cite{ekrt2,dkmn}.

\acknowledgments 
This work was supported by  
the Director, Office of Energy 
Research, Office of High Energy and Nuclear Physics, 
Division of Nuclear Physics, and by the Office of Basic Energy Science, 
Division of Nuclear Science, of  
the U.S. Department of Energy 
under Contract No. DE-AC03-76SF00098 and 
DE-FG-02-93ER-40764. X.-N. Wang was also supported in part by 
NSFC under project 19928511.

\end{multicols}  
\end{document}